%% file: main.tex
\def\year{2021}\relax
\documentclass[letterpaper]{article} 
\usepackage{aaai21}  
\usepackage{times}  
\usepackage{helvet} 
\usepackage{courier}  
\usepackage[hyphens]{url}  
\usepackage{graphicx} 
\usepackage{natbib}  
\usepackage{caption} 
\usepackage{booktabs}
\usepackage{amsmath}
\usepackage{bbold}
\usepackage[ruled]{algorithm2e}
\usepackage{amsfonts}
\usepackage{subfigure}
\usepackage{tabularx,multirow}
\newcolumntype{Y}[1]{>{\hsize=#1\hsize\centering\arraybackslash}X}

\DeclareMathOperator*{\argmax}{\arg\!\max}

\title{Inference-Based Deterministic Messaging For Multi-Agent Communication}
\iftrue
\author{
    Varun Bhatt,
    Michael Buro
}
\affiliations{

    Department of Computing Science \\
    University of Alberta, Canada \\
    vbhatt@ualberta.ca, mburo@ualberta.ca

}
\else
\author{Paper \# 10060}
\fi

\begin{document}

\maketitle

\begin{abstract}
        \input{source/abstract.tex}
    \end{abstract}

    \section{Introduction}

    \input{source/intro.tex}

    \label{sec:intro}

    \section{Background and Related Work}

    \input{source/background.tex}

    \section{Inference-Based Messaging}

    \input{source/method.tex}

    \section{Signaling Game Experiments}

    \input{source/matrix_experiments.tex}

    \section{Gridworld Experiments}

    \input{source/gridworld_experiments.tex}

%

    \section{Conclusions}

    \input{source/conclusions.tex}

    \section*{Acknowledgements}

    \input{source/acknowledgements.tex}

    \small
    \bibliography{ref}

    \clearpage

    \normalsize
    \input{source/appendix.tex}

\end{document}

%% file: source/abstract.tex
Communication is essential for coordination among humans and animals.
Therefore, with the introduction of intelligent agents into the world,
agent-to-agent and agent-to-human communication becomes necessary.  In this
paper, we first study learning in matrix-based signaling games to empirically
show that decentralized methods can converge to a suboptimal policy. We then
propose a modification to the messaging policy, in which the sender
deterministically chooses the best message that helps the receiver to infer
the sender's observation.  Using this modification, we see, empirically, that
the agents converge to the optimal policy in nearly all the runs. We then
apply this method to a partially observable gridworld environment which
requires cooperation between two agents and show that, with appropriate
approximation methods, the proposed sender modification can enhance existing
decentralized training methods for more complex domains as well.

%% file: source/intro.tex
Humans rely extensively on communication to both learn quickly and to act
efficiently in environments in which agents benefit from cooperation. As
artificial intelligence (AI) applications become commonplace in the real
world, intelligent agents therefore can benefit greatly from being able to
communicate with humans and each other. For example, a group of self-driving
cars can improve their driving performance by communicating with other cars
about what they see and what they intend to do~\citep{yang2004vehicle}.  As
advances in other fields of AI have shown, a learned solution is often better
than a manually designed one~\citep{he2015delving,silver2018general}.  Hence,
training the agents to learn to communicate has the potential to lead to more
efficient protocols than pre-defined ones.

One assumption that is commonly held when studying communication between agents
is that messages do not directly affect the payoffs or the rewards that the
agents obtain, which is also known as the ``cheap-talk''
assumption~\citep{crawford1982strategic}. While it does not
accurately reflect all real-world scenarios, it is a reasonable assumption in
many cases. For example, turn indicators and traffic lights do not affect
driving directly because the driving outcome only depends on the drivers'
actions, but not on the state of the lights.

\begin{figure}[b]
  \centering
  \includegraphics[width=0.95\columnwidth]{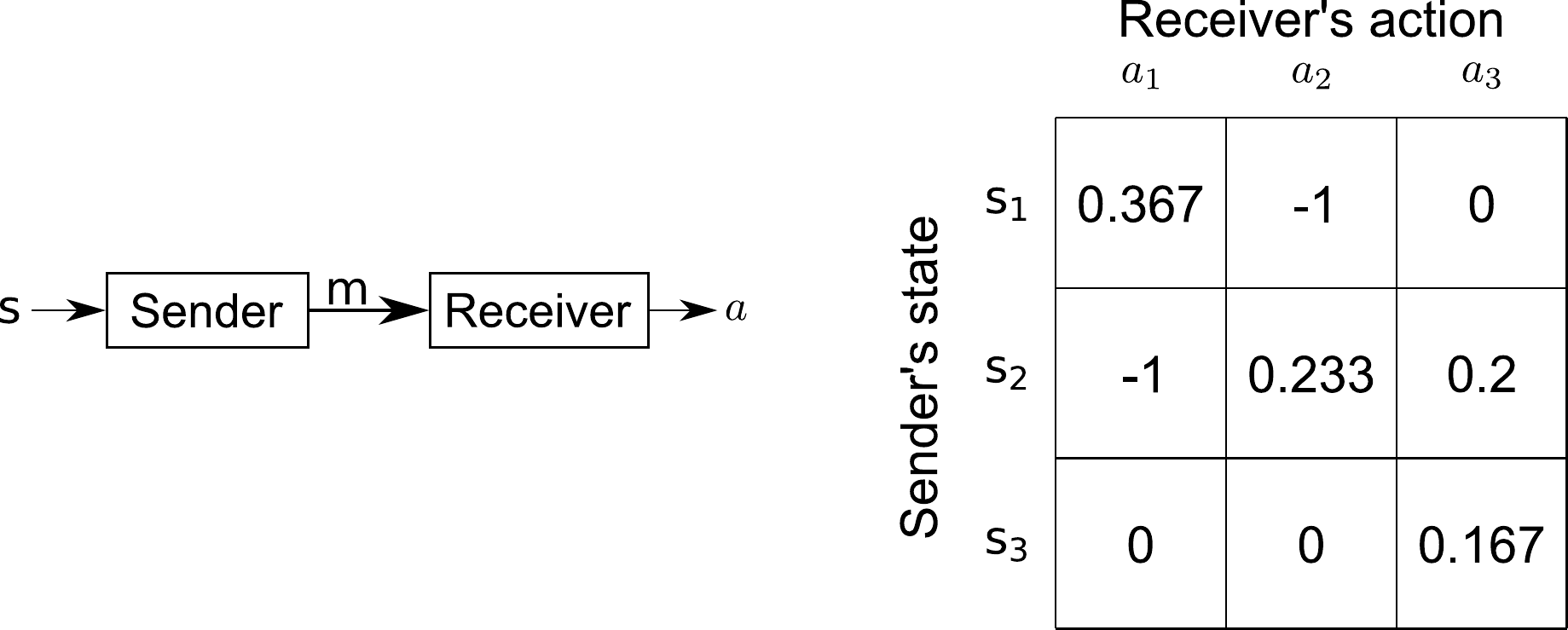}
  \caption{Signaling game with the normalized payoff matrix of the climbing game.}
  \label{fig:climbing_game}
\end{figure}

The aim of this paper is to study the performance of learning algorithms in
synthetic cooperation tasks involving rational agents that know that
they are in a cooperative setting and require communication to act optimally.
First, we consider
one of the simplest setting to study communication: one-step, two-agent
cooperative signaling
games~\citep{gibbons1992primer} depicted in Fig.~\ref{fig:climbing_game}, with
arbitrary payoffs such as those used in the climbing game
(\cite{claus1998dynamics}). In every round of such
games, the sender receives a state $s$ from the environment and then sends
message $m$ to the receiver which, based only on $m$, takes action $a$. In the
problems studied here, both agents receive the same payoff $R(s, a)$,
independent of $m$.
We restrict our studies to decentralized training methods that learn from
experience.
In the algorithms we consider, each agent maintains its own parameters and
updates them only dependent on its private observations and rewards.
Such methods are more scalable compared to centralized training and allow the
agents to keep their individual training methods private while still allowing
cooperation.

We also consider a more complex gridworld cooperation task, in which
two agents need to share their private information using a limited communication
channel while simultaneously acting in the environment to achieve the best
possible performance.
Since both agents receive the same rewards in both
domains, the agents are always motivated to correctly communicate their
private observations.

In what follows, we first discuss related work and then propose a method of
communication based on the sender choosing the best message that would lead to the
correct inference of its private observation.  In the tabular case of a
signaling game, the sender exactly simulates the receiver's inference process,
whereas, in the multi-step gridworld environment, an approximation is
necessary. We then empirically show that finding optimal policies
incrementally through playing experience can be difficult for existing
algorithms even in one-step signaling games, but our method manages to reach
an optimal policy in nearly all the runs. Finally, we present and discuss the
results of our approximation-based method applied to a more complex gridworld
environment.

%


%% file: source/background.tex
%

\subsection{Multi-Agent Reinforcement Learning}

Reinforcement learning (RL)~\citep{sutton2018reinforcement} is a learning
paradigm that is well suited for learning incrementally through experience,
and has been successfully applied to single-agent~\citep{mnih2015human} and
adversarial two-player games~\citep{silver2018general,vinyals2019alphastar}.

A multi-agent reinforcement learning (MARL) problem~\citep{littman1994markov}
consists of an environment and $N\ge 2$ agents, formalized using Markov
games~\citep{shapley1953stochastic}: At each time step $t$, agents receive
state $s^{(t)} \in \mathcal{S}$. Each agent $i$ then takes action $a^{(t)}_{i}
\in \mathcal{A}_i$ and receives reward $r^{(t+1)}_{i}$ and the next state
$s^{(t+1)}$. The distributions of the reward and the next state obey the
Markov property: $p(\mathbf{r^{(t+1)}}, s^{(t+1)}|s^{(\leq t)},
\mathbf{a^{(\leq t)}}) = p(\mathbf{r^{(t+1)}}, s^{(t+1)}|s^{(t)},
\mathbf{a^{(t)}})$. With partial observability or incomplete information,
instead of the complete state, the agents only receive private observation
$o^{(t)}_{i}$. In a pure cooperative setting, the rewards $r^{(t)}_{i}$ agents
receive are equal at every time step.

Scenarios involving multiple learning agents can be very complex because of
non-stationarity, huge policy spaces, and the need for effective
exploration~\citep{hernandez2019survey}.  One way to solve a MARL problem
is to independently train each agent using a single-agent RL algorithm,
treating the other agents as a part of the environment.  However, due to
non-stationarity, the convergence guarantees of the algorithms no longer
exist~\citep{bowling2000analysis}. Additionally, when more than one
equilibrium exists, selecting a Pareto-optimal equilibrium becomes a problem,
in addition to actually converging to
one~\citep{claus1998dynamics,fulda2007predicting}.
WoLF-PHC~\citep{bowling2002multiagent} attempts to solve this problem in
adversarial games, while ``hysteretic
learners''~\citep{matignon2007hysteretic,omidshafiei2017deep} and ``lenient
learners''~\citep{panait2006lenience,palmer2018lenient} are examples of algorithms designed for
convergence to a Pareto-optimal equilibrium in cooperative
games. A comprehensive survey of learning algorithms for multi-agent
cooperative settings is given by
\citet{panait2005cooperative,jan2005overview,busoniu2008comprehensive,tuyls2012multiagent,matignon2012independent,hernandez2019survey}.

\subsection{Multi-Agent Communication}

In the context of MARL problems, communication is added in form of messages that
can be shared among agents.
At each time step $t$, each agent $i$ sends a message $m^{(t)}_{i} \in
\mathcal{M}_i$ in addition to taking an action.
Agents can use messages from other agents to take more informed actions.

Communication in multi-agent systems was initially studied using fixed protocols
to share observations and
experiences~\citep{tan1993multi,balch1994communication}.
It was found that communication speeds up learning and leads to better
performance in certain problems.


In some of the newer studies, such as DIAL~\citep{foerster2016learning} and
CommNet~\citep{sukhbaatar2016learning}, agents are trained by
backpropagating the losses through the communication channel. However, these
methods require centralized training, which may not always be possible in
real-world settings.

On the other hand, \citet{jaques2018intrinsic} and \citet{eccles2019biases} focus on
decentralized training.  \citet{jaques2018intrinsic} use the idea of social
influence to incentivize the sender to send messages that affect the actions
of the receiver.  \citet{eccles2019biases} add additional losses to the agents
in addition to the policy gradient loss.  The sender optimizes an information
maximization based loss while the receiver maximizes the usage of the received
message.

Communication is also studied in the context of emergence of
language~\citep{wagner2003progress,lazaridou2016multi,evtimova2017emergent,lowe2019pitfalls}
in which agents are trained to communicate to solve a particular task with the
goal of analyzing the resulting communication protocols.
One of the commonly
used problem setting is that of a referential game which is a special case of a
signaling game in which the reward
$R(s, a)$ is 1 if $s$ and $a$ match and 0 otherwise.  In this paper, we show
that the problem becomes considerably harder when using an arbitrary payoff
matrix and we propose methods
to overcome this issue.

%% file: source/method.tex

In a multi-agent communication problem, agents could potentially require
access to the private state of other agents to be able to find an optimal
action. By contrast, in a decentralized setting, the only information available
about the private state is through the received message.

One way for the receiver to build beliefs about the private state of the
sender is through Bayesian inference of private states given the
message.
Mathematically, given a message $m$, prior state probabilities $p(s)$ for each
state $s$, and model of the sender messaging policy $p(m|s)$ the posterior
probabilities are given by $p(s|m) = {p(m|s) \cdot p(s)}~/~{\sum_{s'}
  p(m|s')\cdot p(s')}$.
The receiver then uses the posterior belief for its action selection.  For
example, it can assume that the current state of the sender $s^{(t)}$ equals
$\argmax_s p(s|m^{(t)})$ and act accordingly, or maximize the expected
reward w.r.t\ the posterior.  There are two issues with this:
During decentralized training, the receiver does not have access to
the sender's messaging policy, and, even if the receiver accurately
models the sender's messaging policy, the posterior state probabilities would
not be useful if the sender's messaging policy is not good.

In our method, we use the inference process to improve the sender's message.
The sender calculates the posterior probabilities of its current state for each
possible message that it can send. It then chooses the message that leads to the
highest posterior probability. Intuitively, the sender is assuming that the
receiver is performing inference and chooses the message that is most likely
to lead to correct inference. Mathematically, the chosen message $m^{(t)}$ is
given by
\begin{equation}
    \label{eq:policy}
    \begin{aligned}
        m^{(t)} &= \argmax_m p(s^{(t)}|m) \\
        &= \argmax_m {p(m|s^{(t)})}~/~{\sum_{s'} p(m|s')p(s')}
    \end{aligned}
\end{equation}
The $p(s^{(t)})$ term is not present in the numerator because it is constant.
Henceforth, we will use the term \textbf{\emph{unscaled messaging policy}} to refer
to $p(m|s)$ and \textbf{\emph{scaled messaging policy}} to refer to
the above inference-based policy.

$p(s|m)$ is undefined if message $m$ is not used by the sender, i.e., $\forall
s: p(m|s)=0$. While it can be set to an arbitrary value in this case, setting
it to 1 allows the sender to explore unused messages, which potentially leads
to the policy being updated to use such messages.

The unscaled messaging policy $p(m|s)$ can be learned using any RL
algorithm. For example, in our experiments, we use a value-based method,
Q-Learning~\citep{watkins1992q}, for the
matrix signaling games and an asynchronous off-policy policy gradient method,
IMPALA~\citep{espeholt2018impala}, in the gridworld
experiments. While the sender simulates inference, it does not require the
receiver to infer the private state of the sender to work well.  In fact, in our
experiments, the receiver is trained using standard RL algorithms:
Q-Learning and REINFORCE~\citep{williams1992simple} for matrix games, and
IMPALA for gridworld.




Algorithm~\ref{alg:sender} lists the pseudocode of the described sender
algorithm for tabular cases.

\begin{algorithm}[t]
\DontPrintSemicolon
\KwIn{Step size $\alpha$}
Initialize $Q(s,m)$ arbitrarily \;
$\forall s \in \mathcal{S}: N(s) \leftarrow 0$ \;
\For{$t \leftarrow 1, 2, \ldots$}{
Receive state $s$ from the environment \;
$N(s) \leftarrow N(s) + 1$ \;
\For{$s' \in \mathcal{S}$}{
$p(s') \leftarrow {N(s')}~/~{\sum_{s''}N(s'')}$ \;
$\pi(s') \leftarrow \argmax_{m'} Q(s', m')$ \;
}
$\forall m \in \mathcal{M}: p(s|m) \leftarrow \begin{cases}
{1} &~ {\sum_{s'} \mathbb{1}_{\{m=\pi(s')\}}p(s')=0} \\
\frac{\mathbb{1}_{\{m=\pi(s)\}}}{\sum_{s'} \mathbb{1}_{\{m=\pi(s')\}}p(s')} &~ \text{otherwise}
\end{cases}
$\;

$m \leftarrow \argmax_{m'} p(s|m')$ \;
Send $m$ to the receiver \;
Observe $r$ after the receiver acts \;
$Q(s,m) \leftarrow Q(s,m) + \alpha (r - Q(s,m))$ \;
}
\caption{Inference-Based Messaging (Signaling Game, Tabular Case)}
\label{alg:sender}
\end{algorithm}

\subsection{Approximating Posterior Probabilities}

In a small matrix game, the posterior state probability, and consequently, the
message to send, can be calculated exactly by using Eq.~\ref{eq:policy}.
But in more complex environments, in
which the state space is large or even infinite, it is necessary to
approximate the probabilities.
Further, with partial observability and multiple time-steps in the episode, the
state would be replaced by the history of received messages and observations.
In practice, we use an LSTM~\citep{hochreiter1997long} that maintains a summary
of the history in the form of its hidden state, $h$.
The LSTM calculates its current hidden state, $h^{(t)}$, using the current
observation, $o^{(t)}$, the received message, $m_{-i}^{(t-1)}$, and the
previous hidden state, $h^{(t-1)}$, as the input.
This hidden state, $h^{(t)}$, is equivalent to a Markov state in an MDP.
Hence, in this paper, we use the term \textbf{\emph{state}} (or $s$) even in
partially observable problems.

We use a simple empirical averaging method to
approximate $p(s|m)$ while the agents are acting in the environment.
The numerator in Eq.~\ref{eq:policy} can be calculated using the unscaled
messaging policy. The denominator can be written as $p(m) =
\mathbb{E}_s[p(m|s)]$. Since we use online training for the agents, the samples
of states that it receives are according to the state distribution $p(s)$ given
by the combination of the environment dynamics and the current action policies
of the agents. Hence, we can approximate the expectation $\mathbb{E}_s[p(m|s)]$
as the empirical mean of the unscaled messaging probabilities that is calculated
at each time step.

In practice, due to the small number of samples in a rollout batch, the variance
in the mean estimate is high, reducing the quality of the messaging policy.
We empirically found that it is better to use estimates from the previous rollout
batches to reduce the variance despite them being biased due to policy updates
that have happened since then.
Mathematically, we maintain an estimate $\hat{p}(m)$ that we update as an
exponentially weighted moving average of the empirical mean calculated during a
rollout $\bar{p}(m)$ with weight $\mu$.

Another way to estimate $p(m)$ is to train a predictor, using the policy
parameters $\theta$ as the input and the true mean as the training signal.
Preliminary tests in the gridworld environment showed that this is a hard
task, possibly due to a large number of policy parameters, the complex
neural network dynamics, and not having access to the true mean.  Empirically,
the performance was lower when using a predictor compared to using a moving
average.

The parameters of the unscaled messaging probabilities can be updated using
any RL algorithm.  The updates need to account for the fact that we are
updating the unscaled messaging policy (the target policy) while acting using
the scaled messaging policy (the behavior policy).
Since the behavior policy has a
probability of 1 for the taken action and the target policy has a probability
$p(m^{(t)}|s^{(t)})$, the importance weight is given by
$\rho^{(t)} = p(m^{(t)}|s^{(t)})~/~1$.
This is explained in detail in
Appendix~\ref{appendix:polup}. Algorithm~\ref{alg:sender_approx} implements these
ideas for the sender in domains requiring approximation.

\begin{algorithm}[t]
\DontPrintSemicolon
\KwIn{Step size $\alpha$, weight $0<\mu<1$, unscaled messaging policy
$p(m|s;\theta)$}
Initialize $\theta$ arbitrarily \;
$\forall m \in \mathcal{M}: \hat{p}(m) \leftarrow {1}~/~|\mathcal{M}|$ \;
\For{rollout $k \leftarrow 1,2, \ldots$}{
$\forall m \in \mathcal{M}: \bar{p}(m) \leftarrow 0$ \;
\For{$t \leftarrow 1,2, \ldots T$}{
Receive state $s$ from the environment \;
$m \leftarrow \argmax_{m'} \big({p(m'|s;\theta)}~/~{\hat{p}(m')}\big)$ \;
$\bar{p}(m) \leftarrow \bar{p}(m) + p(m|s)~/~T$ \;
Send $m$ to the receiver \;
}
$\forall m\in \mathcal{M}: \hat{p}(m) \leftarrow \mu \hat{p}(m) + (1-\mu) \bar{p}(m)$\;
Update $\theta$ using any RL algorithm after off-policy correction \;
}
\caption{Inference-Based Messaging (Gridworld, Approximation Case)}
\label{alg:sender_approx}
\end{algorithm}

Certain problems require the agents to simultaneously act in the environment and exchange messages.
In such cases, a single agent acts as both the sender and the receiver.
Each agent first calculates a Markov state $s$ (e.g.\ with LSTM) using the
current observation, received message, and the previous state.
At this stage, the agent acts as a receiver.
Each agent also maintains an action policy, $p(a|s)$, and a messaging policy,
$p(m|s)$, which are used to select the action to take in the environment and the
message to send to the other agent respectively.
The agent acts as a sender at this stage.
When using inference-based messaging, the message to be sent is selected
from the scaled messaging policy instead of the unscaled messaging policy.

%% file: source/matrix_experiments.tex
\label{section:matrix_exp}


To show the effectiveness of our proposed methods we conducted three
experiments using signaling games: First, we used the climbing game
(Figure~\ref{fig:climbing_game}) to explain the issue of convergence to
sub-optimal policies.
We then performed experiments on two sets of 1,000 payoff matrices of sizes
$3\!\times\!3$ and $32\!\times\!32$, respectively, with each payoff generated uniformly at random
between 0 and 1 and normalized such that the maximum payoff is 1, to show that
the observed convergence issues are not specific to the climbing game.

Each run of our experiments lasted for 1,000 episodes in $3\!\times\!3$ matrix games and
25,000 episodes in $32\!\times\!32$ matrix games.
Using a higher number of
episodes gave qualitatively similar results and hence, we restricted it to the
given numbers.
In each episode, first, a uniform random
state was generated and passed to the sender. The sender then computed its
message and sent it to the receiver, which used the message to compute its
action. Finally, the reward corresponding to the state and the action was sent
to both agents and used to independently update their policies. The obtained
rewards (normalized by the maximum possible reward) and
the converged policies of the agents were recorded.

\subsection{Benchmark Algorithms}

We considered two variants of our algorithm - \textbf{Info-Q} and
\textbf{Info-Policy}. In both algorithms, the sender uses
Algorithm~\ref{alg:sender}. All Q-values of the sender were initialized
pessimistically to prevent unwanted exploration by the sender. In Info-Q, the
receiver used Q-Learning with optimistically initialized Q-values, while in
Info-policy, the receiver used REINFORCE~\citep{williams1992simple}.

We first compared our method with variants of traditional single-agent RL
algorithms. In \textbf{Independent Q-Learning (IQL)}, both the sender and the
receiver are trained independently using Q-Learning. In \textbf{Iterative
  Learning (IQ)}, the learning is iterative, with the sender updating its
policy for a certain number of episodes (denoted by ``period'') while the
receiver is fixed, followed by the receiver updating its policy while the
sender is fixed, and so on. In
\textbf{Model the sender (ModelS)}, based on the message sent by the sender and the
model of the sender, the receiver performs inference and assumes the true
state to be the one with the highest probability. The action is then chosen
using the (state, action) Q-table.

We then compared algorithms from the literature that have been shown to be
effective in cooperative games with simultaneous actions. In \textbf{Model the
receiver (ModelR)}, the sender calculates the payoffs that would be obtained if the
receiver follows the modeled policy and selects the message that maximizes
this payoff. This algorithm is similar to the one used by
\citet{sen2003towards}, with the difference being that ModelR
uses epsilon-greedy w.r.t. the max Q-values instead of Boltzmann action
selection w.r.t.~expected Q-values since it was found to be better
empirically. In \textbf{Hysteretic-Q}~\citep{matignon2007hysteretic}, higher
step size is used for positive Q-value updates and lower step size for
negative Q-value updates. In \textbf{Lenience}~\citep{panait2006lenience}
(specifically the RL version given by \citet{wei2016lenient}),
only positive updates are made on Q-values during initial time steps, ignoring
low rewards.  As the number of episodes increases, Q-values are always updated.
In \textbf{Comm-Bias}, we used REINFORCE to independently train both the sender
and the receiver. The sender and the receiver additionally use the positive
signaling and the positive listening losses given by \citet{eccles2019biases}
during training.
More information about the benchmark algorithms can be found in
Appendix~\ref{appendix:algorithms}.

\subsection{Results for the Climbing Game}

Experiments with the climbing game were performed using the payoff matrix
given in Figure~\ref{fig:climbing_game} to highlight the issues with the
existing algorithms.
%
%
%

%
%
%
%

\begin{figure}[t]
  \centering
  \includegraphics[width=0.75\columnwidth]{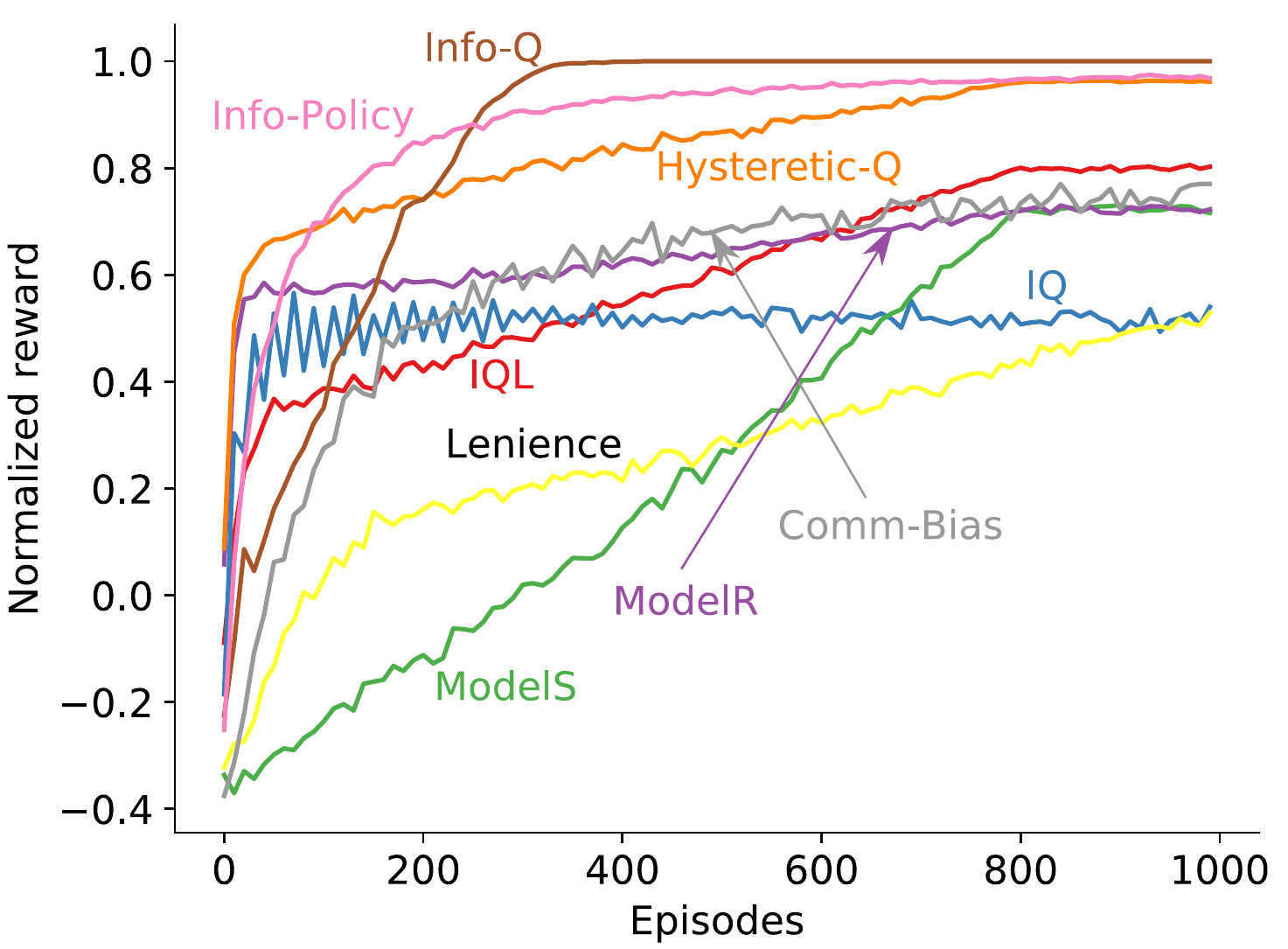}
  \caption{Normalized reward obtained during
    training, as a function of episodes in the climbing game. Info-Q converged to an optimal policy in around
    300 episodes.}
  \label{fig:climbing_rewards}
  \vspace{0.1cm}
  \includegraphics[width=0.75\columnwidth]{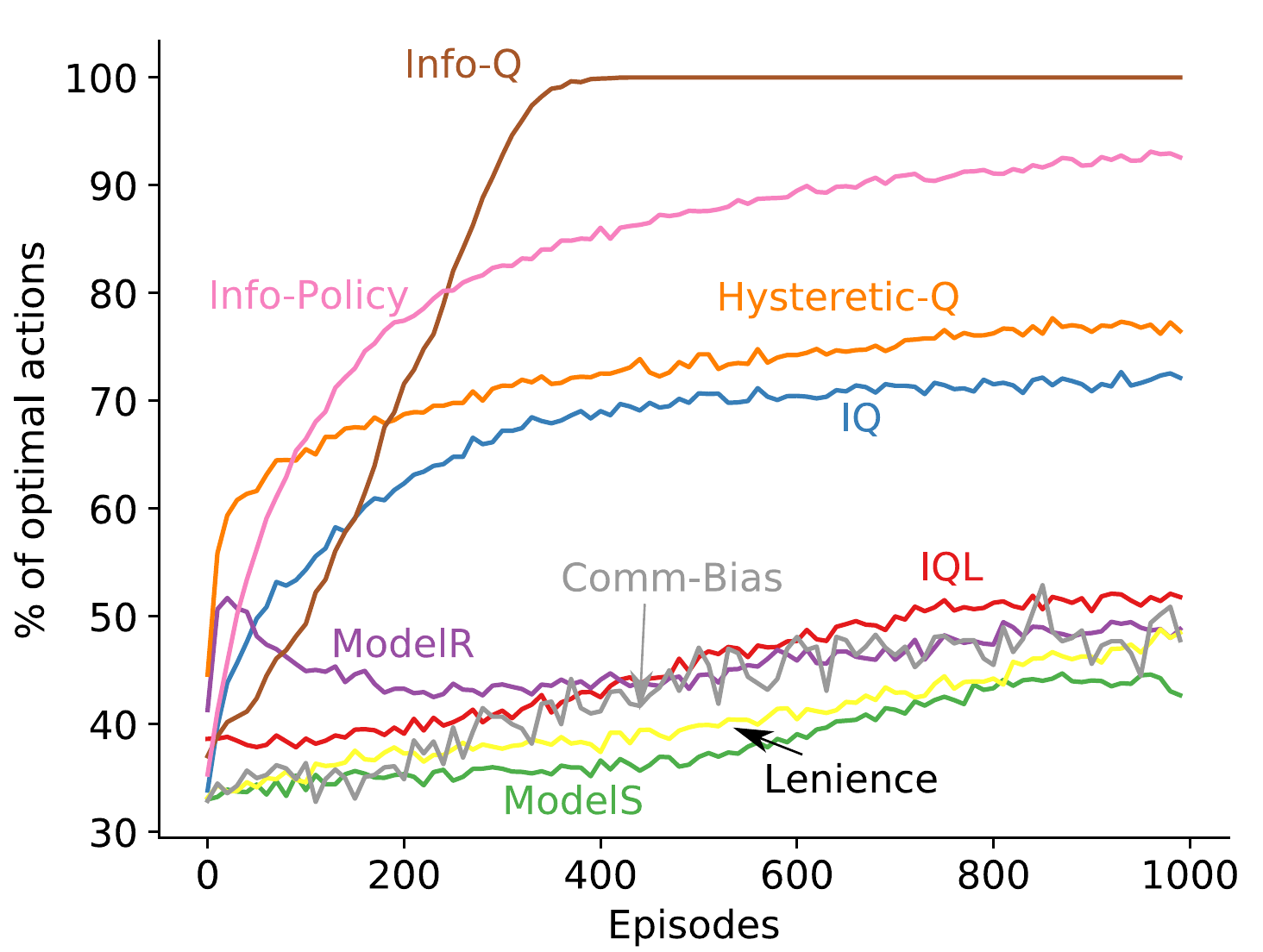}
  \caption{Percentage of optimal actions during training, as a
    function of episodes in the climbing game. The difference between the algorithms is magnified
    in this plot.}
  \label{fig:climbing_opts}
\end{figure}

\begin{figure}[t]
  \centering
  \includegraphics[width=0.9\columnwidth]{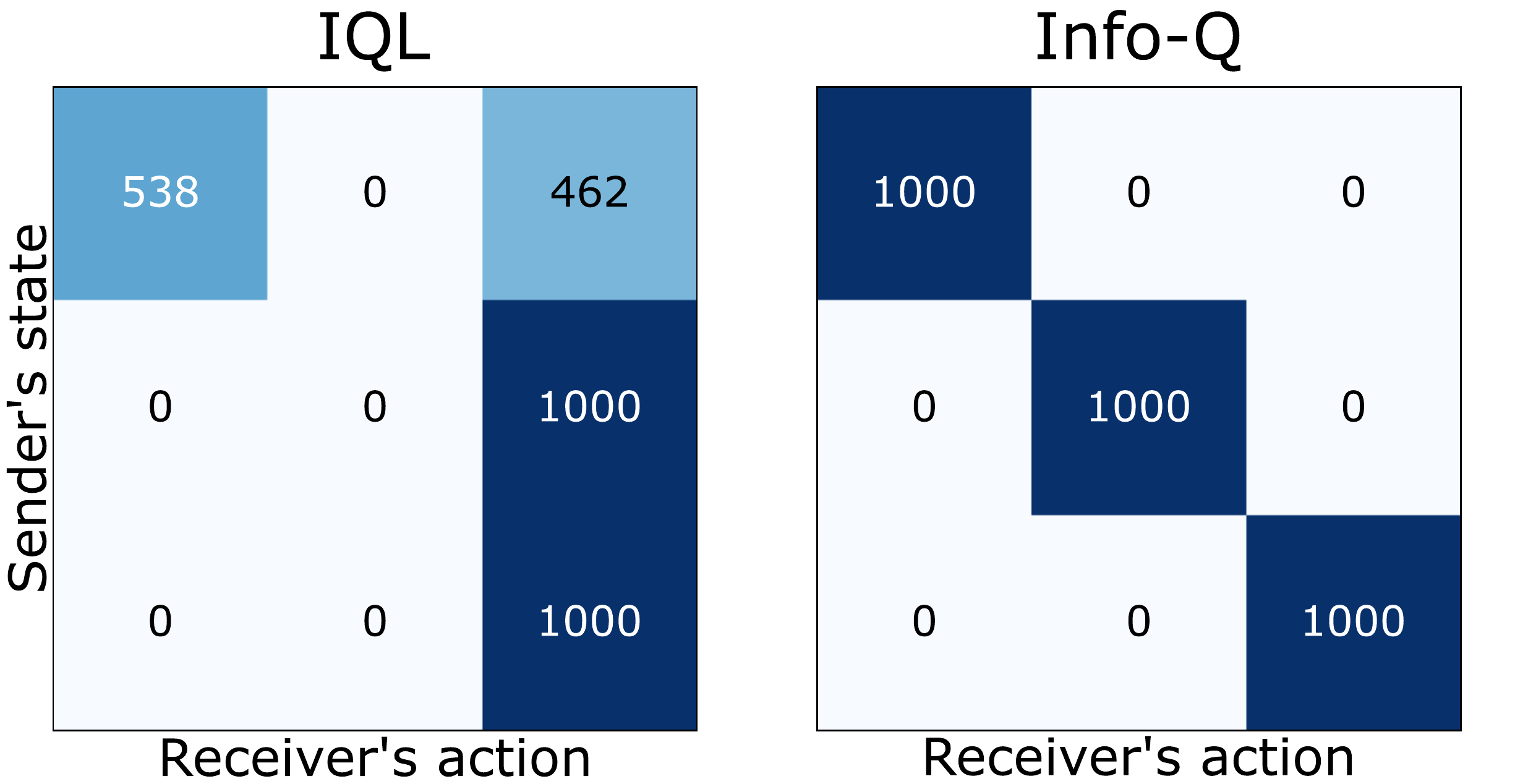}
  \caption{Counts of (state, action) pairs for IQL
    and Info-Q. With IQL, the receiver took $a_3$ in $s_2$, even though $a_2$
    is the optimal action to take. With Info-Q, all runs converged to an
    optimal policy.}
  \label{fig:climbing_cps}
\end{figure}

\begin{figure}[t]
  \centering
  \includegraphics[width=0.8\columnwidth]{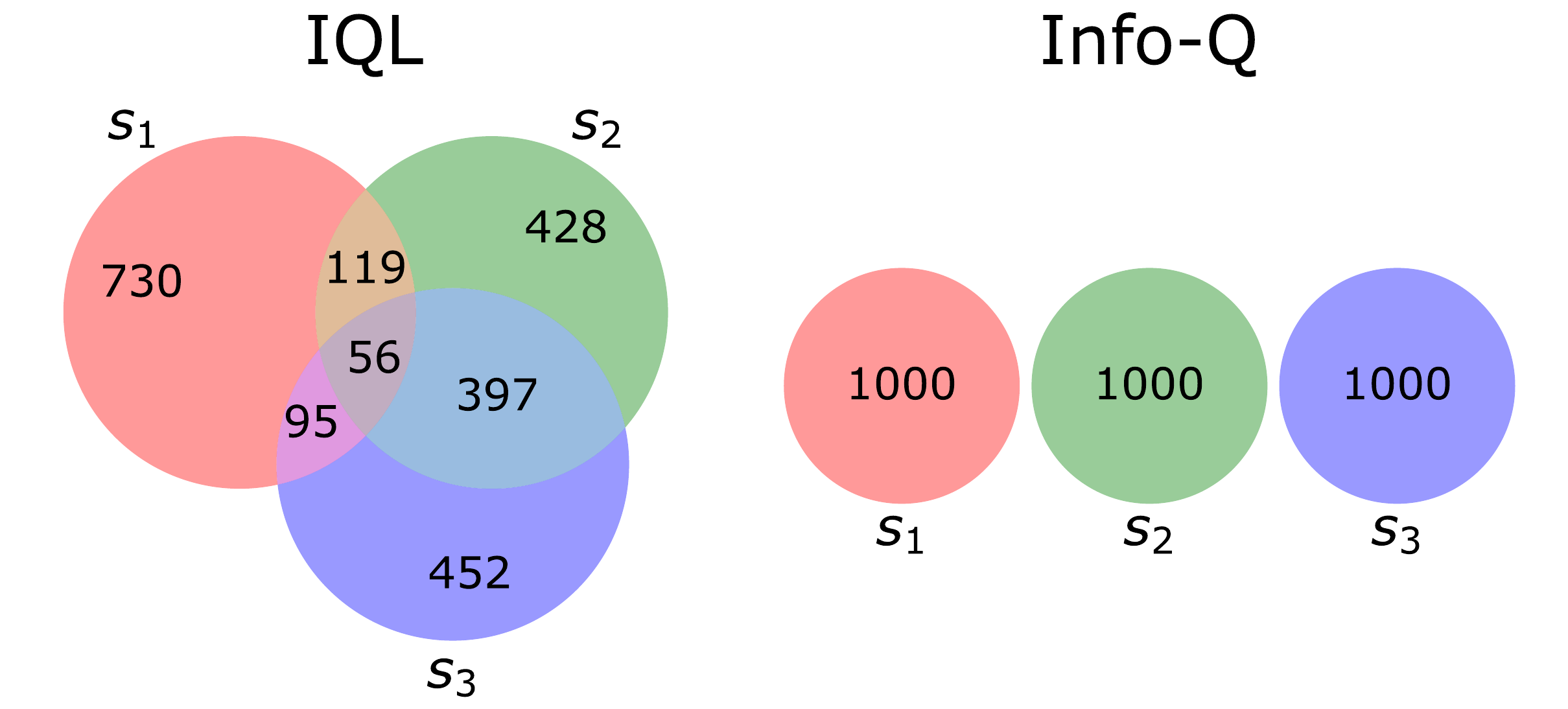}
  \caption{Venn diagram of messaging policy for IQL and
    Info-Q. Intersections in case of IQL imply that in some runs, multiple
    states were assigned the same message. In case of Info-Q, all states were
    assigned a distinct message.}
  \label{fig:climbing_venns}
\end{figure}

Figure~\ref{fig:climbing_rewards} shows the plot of the mean normalized reward, as
a function of episodes. One standard error is shaded, but
less than the line width in most cases. While some of the algorithms obtain rewards close to
that obtained by Info-Q, the difference is magnified in
Figure~\ref{fig:climbing_opts} which shows the percentage of runs that took the
optimal actions, as a function of episodes.

For obtaining more insight into the potential issues of the baseline
algorithms, we counted (state, action) pairs for each run. We iterated through
the states, and for each state, we calculated the corresponding message sent by
the sender and the action taken by the
receiver. Figure~\ref{fig:climbing_cps} shows the matrix with the
counts for each (state, action)-pair in case of Independent Q-Learning (IQL,
on the left) and Info-Q (on the right). It can be observed that in state $s_2$
(middle row), the receiver takes inferior action $a_3$ (last column) with
Q-Learning, whereas Info-Q takes optimal action $a_2$.

We also plotted Venn diagrams to visualize messaging
policies. Figure~\ref{fig:climbing_venns} shows the Venn diagrams for
independent Q-Learning and Info-Q. The region labeled $s_i$ outside of any
intersection corresponds to runs in which $s_i$ was assigned a unique
message. The intersection of regions $s_i$ and $s_j$ denote the runs in which
$s_i$ and $s_j$ were assigned the same message. The intersection of all
regions corresponds to runs in which all states were assigned the same
message. Info-Q assigns a distinct message to each state, whereas
some Q-Learning runs assign the same message to multiple states.



The combination of (state, action) pair counts and the Venn diagrams suggests
that Q-Learning is converging to a sub-optimal policy of choosing $a_3$ in
$s_2$. We hypothesize that this is due to the closeness of the two payoffs and
the fact that $a_3$ is a ``safer'' action to take since the penalty is not
high if the message was incorrect. The messages corresponding to $s_2$ and
$s_3$ are the same in many runs. A possible reason for this is that, since the
receiver is more likely to take $a_3$, there is insufficient incentive to send
distinct messages for $s_2$ and $s_3$. We believe that this vicious cycle
leads to the agents converging to a sub-optimal policy. Info-Q overcomes this
cycle by forcing the sender to fully utilize the communication channel
irrespective of the incentive it receives through the rewards.
In Appendix~\ref{appendix:hard} we present examples of hard 3x3 payoff
matrices that support our intuition.

\begin{figure}[t]
  \centering
  \includegraphics[width=0.75\columnwidth]{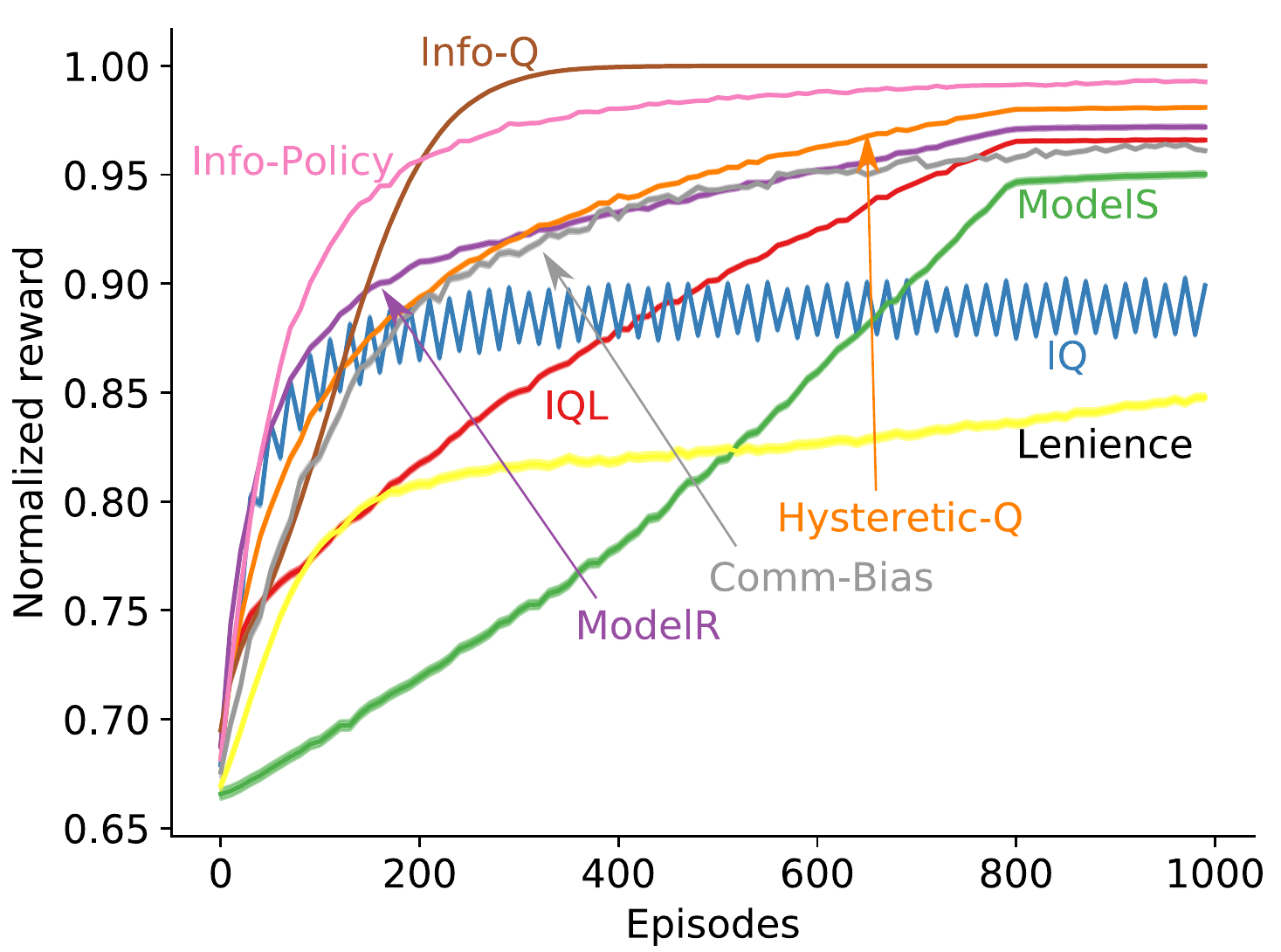}
  \caption{Mean normalized reward obtained during training, as a function of
    episodes, on random $3\!\times\!3$ payoff matrices.}
  \label{fig:random_reward}
\vspace{0.1cm}
  \centering
  \includegraphics[width=0.75\columnwidth]{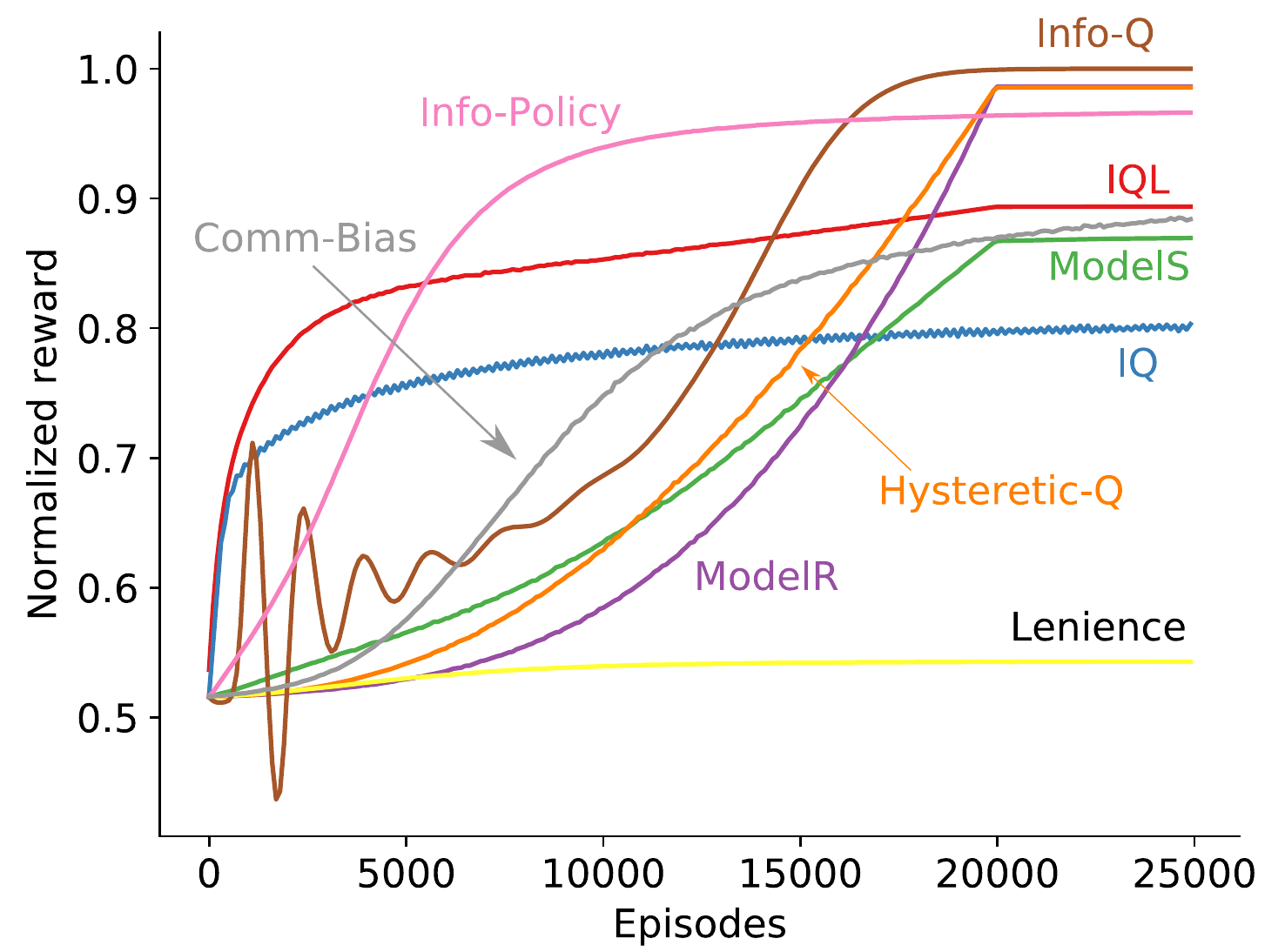}
  \caption{Mean normalized reward obtained during training, as a function of
    episodes, on random $32\!\times\!32$ payoff matrices.}
  \label{fig:random_32x32_reward}
\end{figure}

\subsection{Results for Random Payoff Signaling Games}

\begin{figure}[t]
  \centering
  \includegraphics[width=0.65\columnwidth]{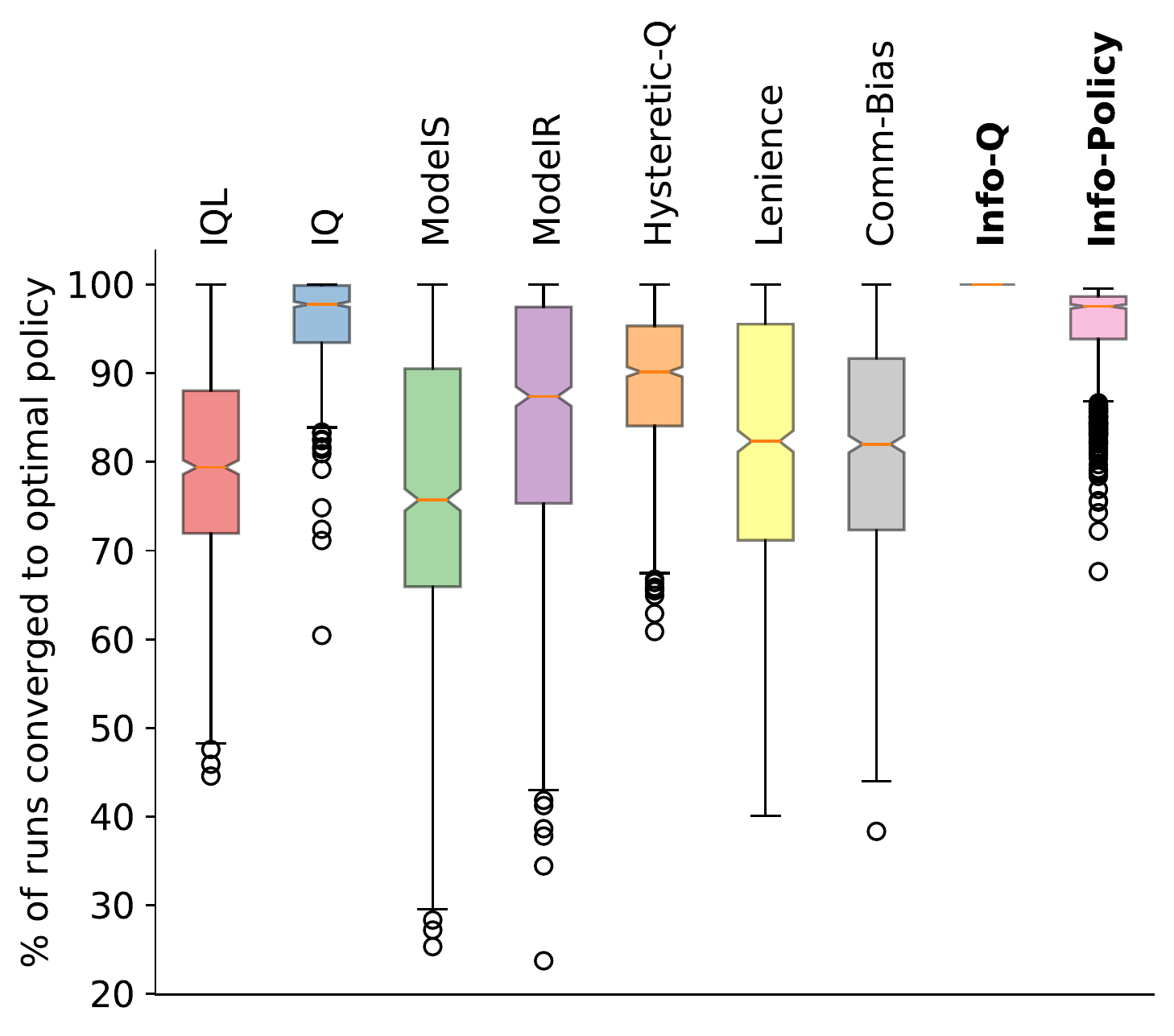}
  \caption{Boxplot of the
    percentage of runs that converged to an optimal policy for each $3\!\times\!3$ payoff
    matrix.}
  \label{fig:random_boxplot}
%
  \centering
  \includegraphics[width=0.65\columnwidth]{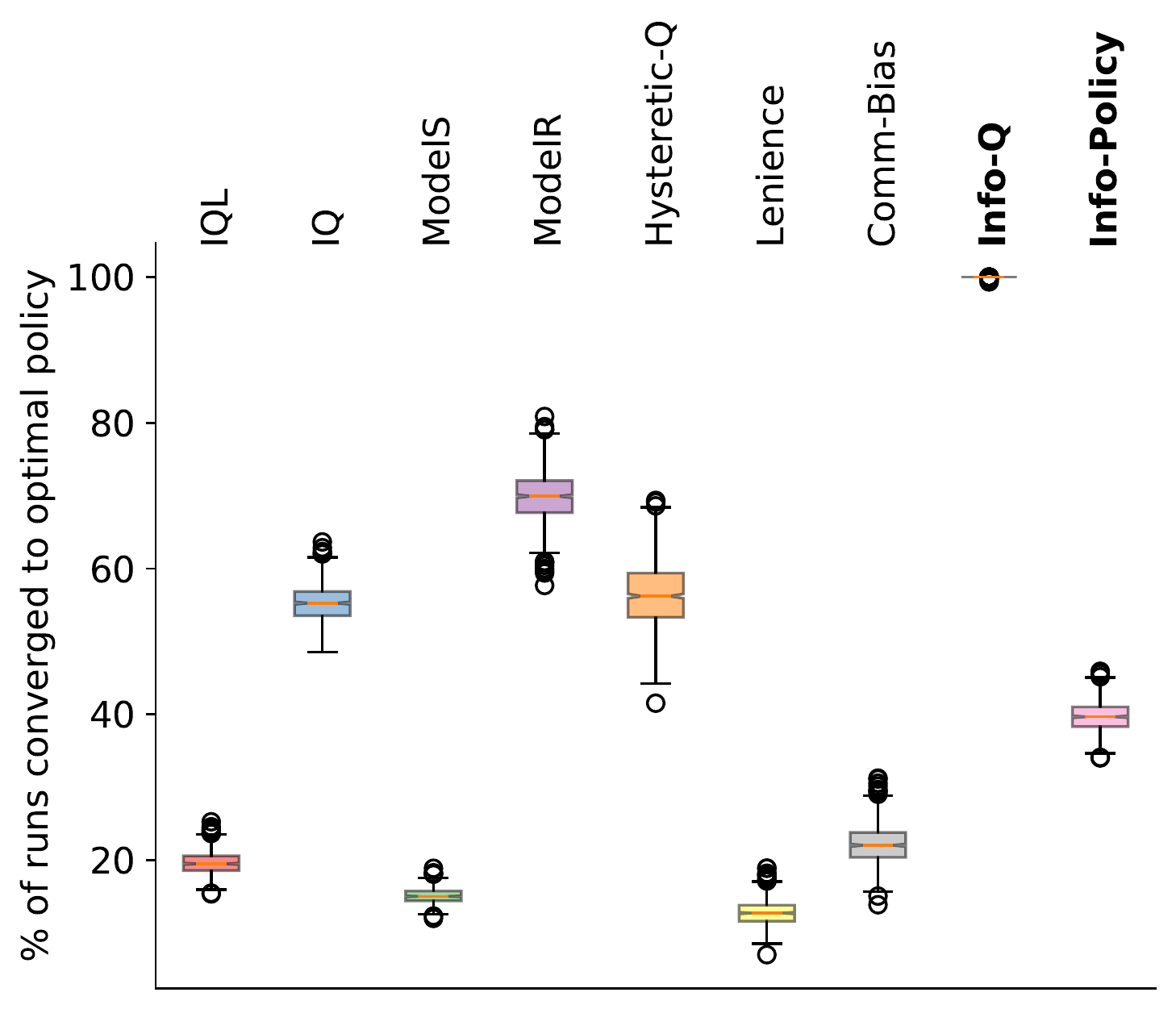}
  \caption{Boxplot of
    the percentage of runs that converged to an optimal policy for each $32\!\times\!32$
    payoff matrix.}
  \label{fig:random_32x32_boxplot}
\end{figure}

To ensure that the issues were not specific to a single game, we next
conducted experiments on randomly generated payoff matrices of size $3\!\times\!3$ and
$32\!\times\!32$.
Figures~\ref{fig:random_reward} and~\ref{fig:random_32x32_reward} show
the plots of the mean normalized reward, as a function of episodes, on $3\!\times\!3$ and
$32\!\times\!32$ payoff matrices respectively.  The mean, here, refers to all random
payoff matrices and multiple runs for each matrix.  One standard error of the mean is
shaded, but less than the line width in most cases.
Figures~\ref{fig:random_boxplot} and~\ref{fig:random_32x32_boxplot}
show the boxplots
of the percentage of runs that converge to an optimal policy for each payoff
matrix of size $3\!\times\!3$ and $32\!\times\!32$ respectively.  The boxplots clearly demonstrate
the advantage of Info-Q compared to the other algorithms in terms of
convergence to an optimal policy.

%

The version of our algorithm with the receiver being trained using policy
gradient (Info-Policy) does not perform as well as Info-Q. This is due to
the policy gradient method itself being slower to learn in this
problem. Comm-Bias is also affected by it and performs worse than
Info-Policy. IQ works fairly well in many problems since it is approximately
equivalent to iterative best response. It fails in cases in which iterative
best response converges to a sub-optimal policy. Since the agents explore more
at the beginning of a period, the obtained reward is much lower. Hence, the
reward curve for IQ does not truly reflect its test-time
performance. Modeling the sender only works well if the receiver has access
to the sender's state during training (but still not as well as
Info-Q). Otherwise, the performance is poor as shown in the
plots. Modeling the receiver suffered from the issue of sub-optimal
policy of the receiver. The best response by the sender to a sub-optimal
receiver policy could be sub-optimal for the problem and vice versa, leading to
the agents not improving.
We postulate that the messages having no effect on the
reward makes it harder for Hysteretic-Q and Lenience. The sender may
receive the same reward for each of the messages it sends, leading to
switching between messages for the same state and confusing the receiver.


%% file: source/gridworld_experiments.tex
\begin{figure}[t]
  \centering
  \includegraphics[width=0.6\columnwidth]{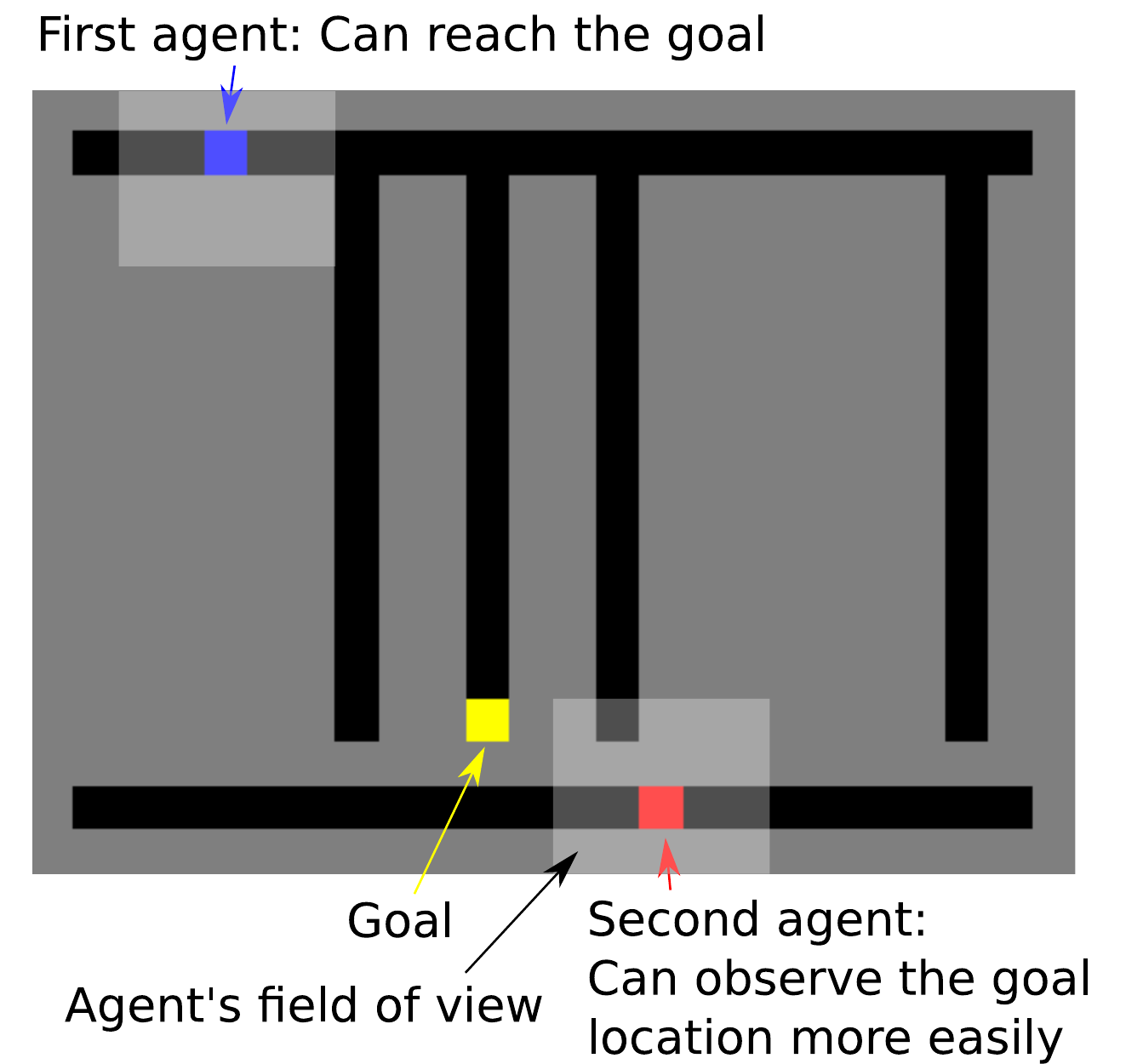}
  \caption{The Treasure Hunt environment.}
  \label{fig:gridworld}
\end{figure}

We use the gridworld environment called ``Treasure Hunt'' introduced by
\citet{eccles2019biases} to test our algorithm in domains in which exact
inference is infeasible (Figure~\ref{fig:gridworld}). There are two agents in
the environment with the goal of reaching the treasure at the bottom of one of
the tunnels.  Both agents have a limited field of view, and their positions
make it so that one agent can easily see the goal location but cannot reach
it, while the other agent can reach the goal, but potentially needs to explore
all tunnels before reaching the goal. By allowing communication, one agent can
easily locate the goal and signal the location to the other agent.
Specifically, both the agents receive the message that the other agent sent at
the previous time step. Hence, both agents take the role of a sender and a
receiver, in contrast to the signaling game, in which there is only one sender
and one receiver.

\begin{table}[t]
    \caption{Comparison of our work with the biases given by \citet{eccles2019biases}}
    \label{tab:comparison}
    \centering
    \small
    \begin{tabularx}{\columnwidth}{Y{1.6}Y{0.5}Y{0.9}}
        \toprule
        \textbf{Method} & \textbf{Final reward} & \textbf{Fraction of good runs} \\
        \midrule
        No-Bias (our implementation) & $11.55 \pm 1.03$ & $0.42$ $(5~/~12)$ \\
        No-Bias~\citep{eccles2019biases} & $12.45 \pm 0.48$ & $0.28$ $(14~/~50)$ \\
        \midrule
        Positive signaling (our implementation) & $16.45 \pm 0.20$ & $1$ $(12~/~12)$ \\
        Positive signaling~\citep{eccles2019biases} & $14.22 \pm 0.36$ & $0.84$ $(42~/~50)$ \\
        \midrule
        Positive signaling + listening (our implementation) & $16.25 \pm 0.20$ & $1$ $(12~/~12)$ \\
        Positive signaling + listening~\citep{eccles2019biases} & $15.14 \pm 0.33$ & $0.94$ $(47~/~50)$ \\
        \midrule
        Inference-Based Messaging & $14.29 \pm 1.26$ & $0.92$ $(11~/~12)$ \\
        \midrule
        Inference-Based Messaging + positive listening & $15.48 \pm 0.76$ & $0.92$ $(11~/~12)$ \\
        \bottomrule
    \end{tabularx}
\end{table}

We kept the core of the training algorithms including the neural network
architecture used by \citet{eccles2019biases} to make results comparable, and
only modified the communication method. The agents choose messages to be sent
by treating them as additional actions. We used a convolutional neural network
followed by densely connected layers for feature extraction. An LSTM
layer~\citep{hochreiter1997long} over the features is used for dealing with
partial observability, whose outputs are fed into linear layers for action and
unscaled message logits and the value estimate. The action policy and the
unscaled message policies of agents is updated using an off-policy policy
gradient algorithm, IMPALA~\citep{espeholt2018impala}.
Contrastive Predictive Coding~\citep{oord2018representation} is used to
improve the LSTM performance. The experiments were run using the
RLLib~\citep{liang2017rllib} library. In our method, the message selection
uses inference simulation as shown in Algorithm~\ref{alg:sender_approx} to compute
$m^{(t)} = \argmax_m p(s^{(t)}|m)$.

As a baseline, we used agents that picked messages based on the learned unscaled
messaging policy and were trained independently using IMPALA
(called \textbf{No-Bias}). We also compared our method with the biases given
by \citet{eccles2019biases}.
With positive signaling bias, the sender was incentivized into sending a message
with higher information, using losses based on mutual information between
private states and messages.
With positive listening bias, the receiver was incentivized to use the received
message for its action selection.
It was achieved by maximizing the divergence between action probabilities
resulting when messages are used and those when the messages are zeroed out.
Each experiment was repeated 12 times and each
run lasted for 300 million time steps. More details about the environment and
the algorithms can be found in Appendix~\ref{appendix:gridworld_experiments}.

\subsection{Results}

Table~\ref{tab:comparison} shows the results we obtained for methods presented
in \citet{eccles2019biases} and our inference-based method. Similar to their
paper, we divide the runs into two categories: good runs, in which the final
reward is greater than 13, and all runs. Good runs require efficient
communication since the max reward achieved without communication is 13 as
shown by \citet{eccles2019biases}. Due to discrepancies in the results of our
implementation of the communication biases and the values reported by
\citet{eccles2019biases}, we provide both the values in the table. We believe
that the difference in the number of repetitions for each experiment and the
number of training steps to be the reason for these discrepancies. Due to
computational constraints, we couldn't perform longer runs or more
repetitions.

Inference-based messaging performed significantly (more than one standard error
of the mean difference) better than No-Bias. Adding
positive listening bias further improved the performance of inference-based
messaging. The performance of our method is similar to the reported value of
positive signaling bias. Adding positive listening bias further improves our
method, reaching a level similar to that of both signaling and listening biases.
Agents receive more than 13 reward per episode after training in 11 of
the 12 runs with inference-based messaging, which, as described earlier, shows
that the agents are learning to communicate in most of the runs.
\citet{eccles2019biases} also show results of the social
influence algorithm~\citep{jaques2018intrinsic} on this environment.
The reported performance of social influence is nearly equal to that achieved by
the no-bias baseline.
Hence, our method significantly outperforms it too.

The results indicate that inference-based messaging can also improve
communication in complex multi-agent environments, and be complementary to
methods that improve the receiver.


%% file: source/conclusions.tex
The contributions of this paper are threefold. First, we demonstrated that
state-of-the-art MARL algorithms often converge to sub-optimal policies in a
signaling game. By analysing random payoff matrices we found that a
sub-optimal payoff being close to the optimal payoff, in combination with the
sub-optimal action having higher average payoffs in other states can lead to
such behaviour.

We then proposed a method in which the sender simulates the receiver's
Bayesian inference of private state given a message to guide its message
selection. Training agents with
this new algorithm led to convergence to the optimal policy in nearly all
runs, for a varied set of payoff matrices. The motivation to use the full
communication channel irrespective of the reward appears to help learning
agents to converge to the optimal policy.

Finally, we applied our method to a more complex gridworld problem which
requires probability approximations for the inference simulation process. In
this domain, too, we could show performance gains.
However, we believe that with more sophisticated inference approximation
techniques, the performance can be further improved.


%% file: source/acknowledgements.tex
This work was funded by The Natural Sciences and Engineering Research Council of
Canada (NSERC).

%% file: source/appendix.tex
\appendix

{\Large\textbf{Appendix: Supplementary Material}}

\section{Policy Updates in Algorithm~\ref{alg:sender_approx}}

Algorithm~\ref{alg:sender_approx} states that the policy parameters $\theta$ can be
updated using any RL algorithm after off-policy correction. We use
IMPALA~\citep{espeholt2018impala} in our experiments. The policy gradient
in the original IMPALA update is given by
\begin{equation*}
  \rho^{(t)} \nabla_\theta \log p(m^{(t)}|s^{(t)};\theta)(r^{(t)} + \gamma
  v^{(t+1)} - V_\omega(s^{(t)}))
\end{equation*}
$\rho^{(t)}$ in the above equations is the importance sampling weight to account
for off-policy asynchronous actors. $v^{(t+1)}$ is the V-trace target computed
similar to the original paper. $r^{(t)}$ is the current reward and
$V_\omega(s^{(t)})$ is the value prediction given by the critic.

Since our agents use deterministic messaging policy while updating the unscaled
messaging policy, the importance weight is given by $\rho^{(t)} =
p(m^{(t)}|s^{(t)})~/~1$. Hence, the policy gradient becomes
\begin{equation*}
  \nabla_\theta p(m^{(t)}|s^{(t)};\theta)(r^{(t)} + \gamma
  v^{(t+1)} - V_\omega(s^{(t)}))
\end{equation*}

\label{appendix:polup}

\section{Benchmark Algorithm Details}

\label{appendix:algorithms}

This section provides more detailed descriptions and parameter choices of the
algorithms we compared our algorithm with in Section~\ref{section:matrix_exp}.
Our selection includes traditional single-agent RL algorithms and algorithms
from the literature that are shown to be effective in cooperative games
with simultaneous actions.
In particular, we consider:

$\bullet$ \textbf{Info-Q} -- The information maximizing sender algorithm
proposed in this paper. The sender uses Algorithm~\ref{alg:sender} and the
receiver uses Q-Learning. All Q-values of the
sender were initialized pessimistically (to -2) to prevent unwanted
exploration by the sender. All Q-values of the receiver were initialized
optimistically (to +2) to allow faster learning, but the same effect can be
obtained by a higher step size. We used step size 0.1 and greedy action
selection for the receiver. We found that $\epsilon$-greedy
exploration by the receiver was not required for the signaling games we chose
in this paper. The initial hyper-parameter guesses led to 100\% convergence to
optimal policy and hence, were not tuned further.

$\bullet$ \textbf{Info-Policy} -- Similar to Info-Q, but the receiver uses a
policy gradient method instead of Q-Learning. Updates to the receiver
are performed using the REINFORCE update rule~\citep{williams1992simple}
with the value of the state as baseline. Q-values of the sender were
initialized pessimistically (to -2). The step size for both the policy
update and the value baseline update was set to 0.5, while the step size
for Q-value updates for the sender was set to 0.05. The receiver's step size was
tuned between $\{0.1, 0.5\}$ and the sender's step size was tuned between
$\{0.5, 0.05\}$.

$\bullet$ \textbf{Fixed messages or fixed actions} -- For each algorithm, we
ran a version in which the policy of either the sender or the receiver was
fixed to an optimal one and only the other agent is trained. With this, the
problem was reduced to a single-agent problem.

$\bullet$ \textbf{Independent Q-Learning (IQL)} -- Both the sender and the
receiver are trained independently using Q-Learning. The sender
maintains a Q-table with entries for each state and each message, while
the receiver maintains entries for each message and each
action. For $3\!\times\!3$ payoffs, $\epsilon$-greedy, with initial $\epsilon$ of 0.3 and linear
decay by $3.75\!\times\!10^{-4}$ per episode, was used for action selection, and step
size ($\alpha$) 0.1 was used for the updates. For $32\!\times\!32$ payoffs, $\epsilon=0.1, \alpha=0.5$
were used, with $\epsilon$ decaying at a rate of $5\!\times\!10^{-6}$ per episode. $\alpha$ and
initial $\epsilon$ were tuned among $\{0.01, 0.05, 0.1, 0.5\}$ and $\{1, 0.5,
0.3, 0.1\}$ respectively.

$\bullet$ \textbf{Iterative Learning (IQ)} -- The learning is iterative, with
the sender updating its policy for a certain number of episodes (denoted
by `period') while the receiver is fixed, followed by the receiver
updating its policy while the sender is fixed, and so on. During each
period, the updates for either the sender or the receiver are exactly the
same as in independent Q-Learning. An agent doesn't explore when its
policy is fixed. When its policy is being trained, the action selection
is $\epsilon$-greedy, with $\epsilon$ set to 1 at the beginning of the
period and linearly decayed by 0.125 after every episode. Step size 0.5
was used for the updates and periods spanned 10 episodes. For $32\!\times\!32$ payoffs,
periods spanned 100 episodes, and $\epsilon$ decay rate was 0.0125. Step size,
initial $\epsilon$, and period were tuned among $\{0.01, 0.05, 0.1, 0.5\}$,
$\{1, 0.5, 0.3, 0.1\}$, and $\{1, 10, 100\}$ respectively.

$\bullet$ \textbf{Model the sender (ModelS)} -- The receiver maintains a
Q-table for each state and each action. It also maintains a Q-table for each
state and each message to model the sender. Based on the message sent by the
sender and the model of the sender, the receiver calculates the posterior
probabilities of each state ($p(s|m)$). The receiver assumes the true state to
be the one with the highest probability, i.e., $\argmax_s p(s|m)$, and the
Q-table for states and actions is used to compute the best action as
$\argmax_a Q(s, a)$.  $\epsilon$-greedy, with initial
$\epsilon$ of 1 and linear decay by $1.25\!\times\!10^{-3}$ per episode
was used for action selection, and step size 0.05 was used for all the
updates. The $\epsilon$ decay rate was lowered to $5\!\times\!10^{-5}$, while the step size
was increased to 0.1 in the case of $32\!\times\!32$ payoffs. Step size and
initial $\epsilon$ were tuned among $\{0.01, 0.05, 0.1, 0.5\}$ and $\{1, 0.5,
0.3, 0.1\}$ respectively.

$\bullet$ \textbf{Model the receiver (ModelR)} -- The sender maintains a
Q-table for each state and each action. It also maintains a Q-table for each
message and each action to model the receiver. The sender calculates the
payoffs that would be obtained if the receiver follows the modeled policy and
selects the message that maximizes this payoff (with $\epsilon$-greedy for
exploration). This algorithm is similar to the one used by
\citet{sen2003towards}, with the difference being that ModelR
uses $\epsilon$-greedy w.r.t.\ max of Q-values instead of Boltzmann action
selection w.r.t.\ expected Q-values since we empirically found that
$\epsilon$-greedy performed better. An initial $\epsilon$ of 0.1 with linear
decay of $1.25\!\times\!10^{-4}$ per episode was used for action selection, and step size 0.5
was used for all updates. For $32\!\times\!32$ payoffs, $\epsilon=1$, with a decay of
$5\!\times\!10^{-5}$ per episode was used. Step size and
initial $\epsilon$ were tuned among $\{0.01, 0.05, 0.1, 0.5\}$ and $\{1, 0.5,
0.3, 0.1\}$ respectively.

$\bullet$ \textbf{Hysteretic-Q} -- An adaptation of the algorithm given by
\citet{matignon2007hysteretic} to our problems. The idea is to use higher
step size for positive updates and lower step size for negative
updates. $\epsilon$-greedy was used for exploration instead of Boltzmann
exploration since we empirically found that $\epsilon$-greedy performed better
for this problem. $\epsilon$ was initially set to 0.1 and linearly decayed by
$1.25\!\times\!10^{-4}$ per episode. $\alpha=0.5$, $\beta=0.05$ were used as the step sizes
for positive and negative updates respectively. Problems with $32\!\times\!32$ payoffs
used $\epsilon=1$, which decayed by $5\!\times\!10^{-5}$ per episode. $\alpha$ and
initial $\epsilon$ were tuned among $\{0.01, 0.05, 0.1, 0.5\}$ and $\{1, 0.5,
0.3, 0.1\}$ respectively, while $\beta$ was tuned among $\{0.1, 1, 10\}$ times $\alpha$.

$\bullet$ \textbf{Lenience} -- An adaptation of the algorithm presented in
\citet{panait2006lenience} (specifically the reinforcement learning version
given by \citet{wei2016lenient}). In the initial time steps, Q-values are
updated only if the update is positive, ignoring low rewards. As the number of
episodes increases, Q-values are always updated. Step size 0.1 was used for
updates. Boltzmann action selection, with a maximum temperature of 5, a minimum
temperature of $1.6\!\times\!10^{-3}$, and exponential temperature decay ($\delta$) by 0.99 per episode
was used. The action selection moderation factor ($\omega$) and lenience
moderation factor ($\theta$) were set to 0.1 and 1, respectively. The minimum
temperature was lowered to 0 and $\theta$ was increased to 10 in case of $32\!\times\!32$
payoffs. Step size, $\delta$, maximum temperature, $\omega$ and $\theta$ were
tuned among $\{0.01, 0.05, 0.1, 0.5\}$, $\{0.999, 0.995, 0.99\}$, $\{5, 50, 500,
5000\}$, $\{0.1, 1, 10\}$, $\{0.1, 1, 10\}$.

$\bullet$ \textbf{Comm-Bias} -- Both the sender and the receiver are
independently trained using REINFORCE.
The sender additionally uses the positive signaling loss given by
\citet{eccles2019biases} during training.
The auxiliary loss maximizes the mutual information between the sent message and
the current state.
We found that the receiver using the positive listening bias given by
\citet{eccles2019biases} didn't improve the performance in these games.
Since our experiments are in a tabular setting, we calculated the loss exactly
at each timestep instead of averaging over a batch.
A step size of 0.5 was used for policy updates.
Positive signaling loss was given a weight of 0.01.
The weight for average message entropy term ($\lambda$) was set to 0.1 in the
$3\!\times\!3$ case and 0.3 in the $32\!\times\!32$ case, while the entropy target was set to 0.5 in
the $3\!\times\!3$ case and 0 in the $32\!\times\!32$ case. Step size, positive
signaling loss coefficient, $\lambda$, entropy target were tuned among $\{0.1,
0.5\}$, $\{0.001, 0.01, 0.1\}$, $\{0.1, 0.3, 1.0\}$, $\{0.0, 0.5, 1.0, 1.5\}$.

We selected hyper-parameters using a combination of grid search and
manual tuning. We generated 100 random payoff matrices, and for each set of
hyper-parameters and each payoff matrix, we repeated the experiment 1,000 times.
The set of hyper-parameters with the highest percentage of runs that
converged to the optimal policy was used in all other experiments.
Hyper-parameter tuning was performed separately for 3x3 payoff matrices and
32x32 payoff matrices.

We generated 1,000 random payoff matrices to compare the algorithms.
The best set of hyper-parameters was used for each algorithm, and the experiment
was repeated 1,000 times.
The mean of the normalized obtained rewards and percentage of the runs which
converged to the optimal policy were recorded.

\section{Hard 3x3 Signaling Game Instances}

\label{appendix:hard}

\begin{figure*}[t]
  \centering
  \includegraphics[width=0.6\textwidth]{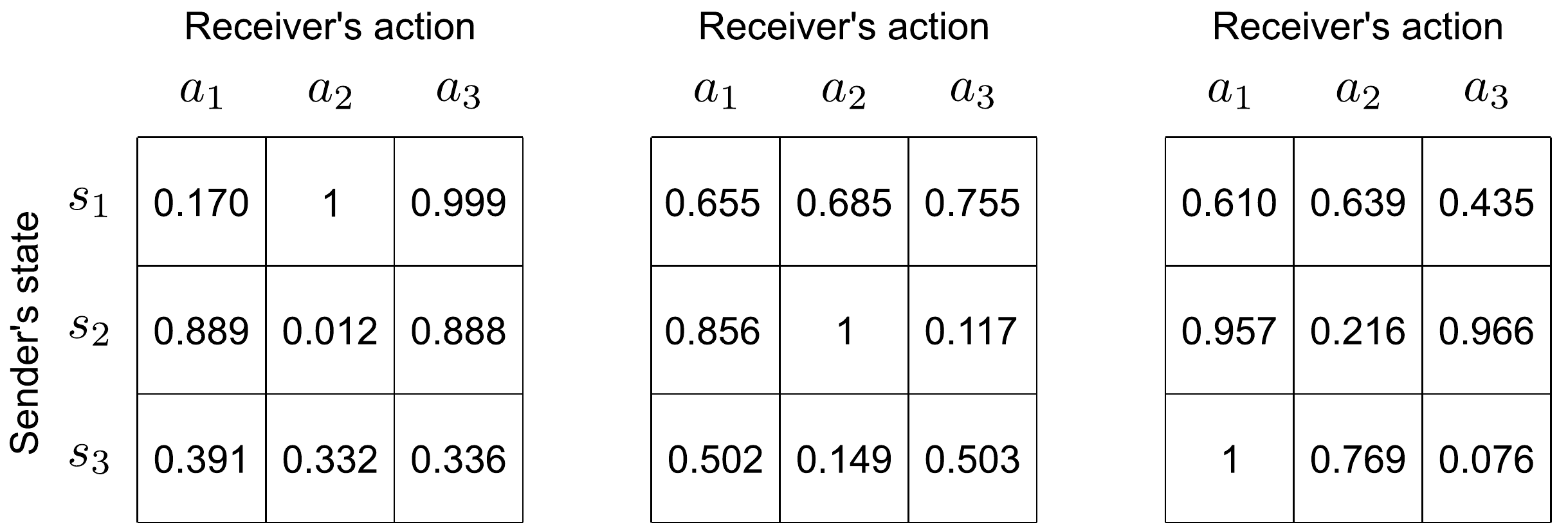}
  \caption{Three payoff matrices among the randomly generated ones that
    resulted in the lowest convergence-to-optimal percentage in terms of the
    mean over all algorithms.}
  \label{fig:hard_payoffs}
\vspace{0.7cm}
  \centering
  \includegraphics[width=0.6\textwidth]{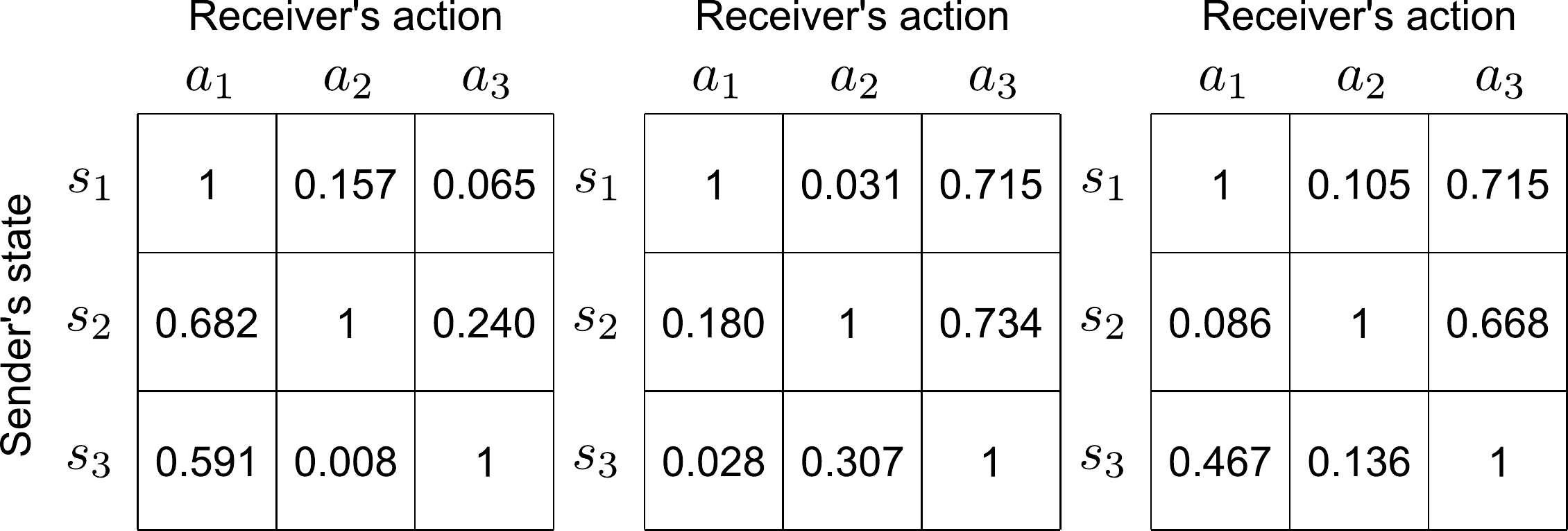}
  \caption{Three payoff matrices among the randomly generated ones that
    resulted in the lowest final mean reward over all algorithms.}
  \label{fig:hard_payoffs_reward}
\end{figure*}

Figure~\ref{fig:hard_payoffs} shows examples of some of the randomly generated
$3\!\times\!3$ payoff matrices that were hard for the algorithms in terms of convergence
to an optimal policy. The three matrices shown in the figure have the
lowest mean percentage of convergence to an optimal policy across all the
algorithms. Info-Q converges to an optimal policy in all the runs with these
payoff matrices. Earlier, we hypothesized that the low percentage of
convergence to optimal policy is due to the closeness in the payoffs of
optimal and sub-optimal actions in a state, combined with a sub-optimal action
having higher payoffs in other states. This hypothesis is strengthened by
these payoff matrices since they too have the same issues. Multiple pairs of
actions in the top game, $a_1$ and $a_3$ in $s_3$ in the middle game, and
$a_1$ and $a_3$ in $s_2$ in the bottom game have close payoffs. In each case,
the sub-optimal action has a higher expected payoff across all the states.

Since the sub-optimal payoffs are close to the optimal payoffs in these
matrices, the algorithms still receive a high reward even when they are choosing
a sub-optimal action.
In contrast, the payoff matrices shown in Figure~\ref{fig:hard_payoffs_reward}
lead to the algorithms receiving the lowest mean reward.
While the percentage of runs that converge to the optimal policy is higher for
these matrices, sub-optimal actions give much lower reward compared to the
optimal ones.
The mean final reward over all algorithms and all runs was around 0.9 for these
matrices.

\section{Gridworld Experiment Details}
\label{appendix:gridworld_experiments}

\subsection{Environment}

The gridworld environment called ``Treasure Hunt'' is used in our experiments.
The environment is kept exactly the same as that defined in the paper
by \citet{eccles2019biases} to keep the comparisons fair.
For each training episode, an 18 high and 24 wide grid is created as below:
\begin{itemize}
  \item Create two tunnels between the second column and the second to last
  column at the second and the second to last rows.
  \item Randomly choose four columns with a spacing of at least two columns
  between each of them.
  Create a tunnel of length 14 starting from the top tunnel in each of these
  columns.
  \item Place the first agent in a random cell in the top horizontal tunnel.
  Place the second agent in a random cell in the bottom horizontal tunnel.
  \item Place the goal at the bottommost cell of a random vertical tunnel.
\end{itemize}

The goal location is moved to the bottommost cell of a random vertical tunnel
every time an agent reaches it.
The episode terminates after 500 time steps and the grid is regenerated using
the above rules.

At each time step, both agents observe a 5x5 area centered around their current
location.
The agents can move in any cardinal direction or take no action.
The agents can also send a message that is shared with the other agent in the
next time step.
If an agent moves into a wall, it stays in its previous location.
If it moves into the goal location, both agents get a reward of 1.

\subsection{Training method}

We trained agents using RLLib~\citep{liang2017rllib} library.
The agents were trained in a decentralized manner using a modified version of
RLLib's implementation of the IMPALA algorithm.
Both agents used their own neural network consisting of policy and value heads
with the following architecture:
\begin{itemize}
  \item Pass the $5\!\times\!5$ observations through a convolutional neural network (CNN)
  with 6 channels, kernel size of 1, and stride of 1.
  \item Flatten the CNN output and concatenate the received message.
  Pass it through two fully connected layers (MLP) with 64 units each.
  \item Concatenate the previous reward and action to the MLP output and pass
  it to an LSTM with 128 hidden units.
  \item Pass the LSTM outputs through a separate linear layer for action
  logits, message logits, and the value estimate.
\end{itemize}

Additionally, Contrastive Predictive Coding (CPC)~\citep{oord2018representation}
objective was added to improve the LSTM performance.
The CPC loss was calculated
as follows:
\begin{itemize}
  \item Transform the LSTM inputs by passing it through a linear layer with 64
  outputs to get the input projection.
  \item Get the output projection corresponding to predictions of inputs at 20
  future time steps by passing the LSTM outputs through 20 different linear
  layers with 64 outputs each.
  \item Compute the dot product of the output projection at time $t$ and input
  projection at time $t+k$ across all batches for $k=1,2,\ldots,20$.
  \item Calculate the CPC loss as the cross-entropy loss using the dot
  products as logits and the prediction being true when the input and output
  projections are from the same batch.
\end{itemize}

The hyper-parameters were selected using those used by
\citet{eccles2019biases} as the starting point and performing grid
  search over the hyper-parameters listed below. The
  other hyper-parameters were unchanged or manually chosen such that the mean
  episode reward after 100 million time steps was maximized.  Each experiment
was repeated 12 times and each run lasted for 300 million time steps.  RMSProp
optimizer~\citep{hinton2012neural} was used for updating the weights, with an
initial learning rate of $10^{-3}$, exponentially decayed by 0.99 after every
million steps.  $\epsilon$ for the optimizer was set to $10^{-6}$. The
optimizer was tuned between Adam~\citep{kingma2014adam} with a learning rate
of $10^{-4}$ and no decay, Adam with a learning rate of $10^{-4}$ and
exponential decay by 0.99 after every million steps, RMSProp with a learning
rate of $10^{-4}$ and no decay, RMSProp with a learning rate of $10^{-3}$ and
exponential decay by 0.99 after every million steps. RMSProp $\epsilon$ was
further tuned between $\{10^{-6}, 10^{-3}\}$.  All gradients were scaled such
that the gradient norm was at most 10. The maximum gradient norm was tuned
between $\{10, 40\}$.  Each rollout consisted of 100 time-steps and a batch
size of 16 was used.  32 asynchronous parallel actors were used to collect
data from the environment.  The batch size was tuned between $\{16, 32\}$, but
rollout length was not tuned.  Policy, value, and entropy losses were balanced
by using a coefficient of 0.5 for the value loss and 0.006 for the entropy
loss.  The value loss coefficient was tuned between $\{0.5, 1\}$, while the
entropy loss coefficient was tuned among $\{0.01, 0.006, 0.001\}$.  The
discount factor was set to 0.99.  The hyper-parameters for the positive
signaling and positive listening biases were the same as those given by
\citet{eccles2019biases}.  For our method, the hyperparameter $\mu$ in
Algorithm~\ref{alg:sender_approx} was set
to 0.5 after tuning among $\{0, 0.25, 0.5, 0.9\}$.

The positive signaling and positive listening biases were implemented
exactly as given by \citet{eccles2019biases}.
The positive signaling loss is given by
\begin{equation*}
  \mathbb{E}(\lambda \mathcal{H}(p(m)) - (\mathcal{H}(m|s) - \mathcal{H}_{target})^2)
\end{equation*}
in which $\lambda$ and $\mathcal{H}_{target}$ are hyper-parameters, while $p(m)$
is calculated using the empirical average of the messaging policy probabilities
over a rollout.

Positive listening loss is calculated in two steps.
First, the policy of an agent without using messages from the other agent
($p(m|s')$) is estimated by passing the observations, with the message from the
other agent is set to zero (resulting in $s'$), through the agent's policy
network to get the estimated policy $\hat{p}(m|s')$.
This estimate is trained by minimizing the loss
\begin{equation*}
  -\sum_{a} p(a|s)\log(\hat{p}(a|s'))
\end{equation*}
by only backpropagating through the $\hat{p}(a|s')$ term.

The distance between the policy which uses messages and the policy which doesn't
is maximized by minimizing the loss
\begin{equation*}
  -\sum_{a} |p(a|s) - \hat{p}(a|s'))|
\end{equation*}
by only backpropagating through the $p(a|s)$ term.

The hyper-parameters used to balance these losses with the policy, value, and
entropy losses, as well as those used in the calculation of the positive
signaling loss are kept the same as those given by \citet{eccles2019biases} for
fair comparison.